\documentclass[aps,pra,preprint,tightenlines,showpacs,groupedaddress]{revtex4}
\usepackage{graphicx}
\usepackage{amsmath}
\usepackage{longtable}

\begin{document}

\title{Large dimension Configuration Interaction calculations of positron binding to
       the group II atoms}
\author{M.W.J.Bromley}
\email{mbromley@physics.sdsu.edu}
\affiliation{Department of Physics, San Diego State University, San Diego CA 92182, USA}
\author{J.Mitroy}
\email{jxm107@rsphysse.anu.edu.au}
\affiliation{Faculty of Technology, Charles Darwin University, Darwin NT 0909, Australia}

\date{\today}

\begin{abstract}

The Configuration Interaction (CI) method is applied to the 
calculation of the structures of a number of positron binding 
systems, including $e^+$Be, $e^+$Mg, $e^+$Ca and $e^+$Sr.  
These calculations were carried out in orbital spaces 
containing about 200 electron and 200 positron orbitals up 
to $\ell = 12$.  Despite the very large dimensions, the 
binding energy and annihilation rate converge slowly 
with $\ell$, and the final values do contain an appreciable    
correction obtained by extrapolating the calculation to the 
$\ell \to \infty$ limit.   The binding energies were 
0.00317 hartree for $e^+$Be, 0.0170 hartree for $e^+$Mg, 
0.0189 hartree for $e^+$Ca, and 0.0131 hartree for $e^+$Sr.  

\end{abstract}


\maketitle

\section{Introduction}

The ability of positrons to bind to a number of atoms is now well 
established \cite{mitroy02b,schrader01a,strasburger03a}, and all of 
the group II elements of the periodic table are expected to bind a 
positron \cite{mitroy02b,mitroy02f}.  There have been two sets of 
calculations that are consistent, in that they tend to predict the 
same binding energy and annihilation rate.  The first set of 
calculations were those undertaken on $e^+$Be and $e^+$Mg 
\cite{ryzhikh98c,ryzhikh98e,mitroy01c} with the fixed core 
stochastic variational method (FCSVM) \cite{ryzhikh98b,ryzhikh98e,mitroy02b}.  
Some time later, configuration interaction (CI) calculations 
were undertaken on $e^+$Be, $e^+$Mg, $e^+$Ca and $e^+$Sr 
\cite{bromley02a,bromley02b}.  The calculations for $e^+$Be 
and $e^+$Mg  agreed to within the respective computational 
uncertainties, which were roughly about 5-10$\%$ for the binding 
energy.   

One feature common to all the CI calculations is the 
slow convergence of the binding energy and the annihilation
rate.   The attractive electron-positron interaction leads 
to the formation of a Ps cluster (i.e. something akin to a 
positronium atom) in the outer valence region of the atom 
\cite{ryzhikh98e,dzuba99,mitroy02b,saito03a}.   The accurate 
representation of a Ps cluster using only single particle 
orbitals centered on the nucleus requires the inclusion of 
orbitals with much higher angular momenta than a roughly 
equivalent electron-only calculation 
\cite{strasburger95,schrader98,mitroy99c,dzuba99}.  For example, 
the largest CI calculations on the group II positronic atoms 
and PsH have typically have involved single particles bases 
with 8 radial function per angular momenta, $\ell$, and 
inclusion of angular momenta up to $L_{\rm max} = 10$ 
\cite{bromley02a,bromley02b,saito03a}.  Even with such large 
orbital basis sets, between 5-60$\%$ of the binding energy and some 
30-80$\%$ of the annihilation rate were obtained by extrapolating 
from $L_{\rm max} = 10$ to the $L_{\rm max} = \infty$ limit.
 
Since our initial CI calculations \cite{bromley00a,bromley02a,bromley02b}, 
advances in computer hardware mean larger dimension CI calculations 
are possible.  In addition, program improvements have removed the 
chief memory bottleneck that previously constrained the size of 
the calculation.  As a result, it is now appropriate to revisit 
the group II atoms to obtain improved estimates of their positron 
binding energies and other expectation values. The new calculations 
that we have performed have orbital spaces more than twice as large
as those reported previously.  The estimated CI binding energies
for all systems have increased, and furthermore the uncertainties  
resulting from the partial wave extrapolation have decreased.

\section{Calculation Method}

The CI method as applied to atomic systems with two valence 
electrons and a positron has been discussed previously  
\cite{bromley02a,bromley02b}, and only a brief description is 
given here.  All calculations were done in the fixed core 
approximation.  The effective Hamiltonian for the system 
with $N_e = 2$ valence electrons and a positron was 

\begin{eqnarray}
H  &=&  - \frac{1}{2}\nabla_{0}^2 - \sum_{i=1}^{N_e} \frac {1}{2} \nabla_{i}^2
- V_{\rm dir}({\bf r}_0) + V_{p1}({\bf r}_0) \nonumber \\
&+& \sum_{i=1}^{N_e} (V_{\rm dir}({\bf r}_i) + V_{\rm exc}({\bf r}_i) + V_{p1}({\bf r}_i)) - \sum_{i=1}^{N_e} \frac{1}{r_{i0}} \nonumber \\
   &+& \sum_{i<j}^{N_e} \frac{1}{r_{ij}}
   - \sum_{i<j}^{N_e} V_{p2}({\bf r}_i,{\bf r}_j)
   + \sum_{i=1}^{N_e} V_{p2}({\bf r}_i,{\bf r}_0) \ .
\end{eqnarray}
The index $0$ denotes the positron, while $i$ and $j$ denote the electrons.
The direct potential ($V_{\rm dir}$) represents the interaction with 
the electronic core, which was derived from a Hartree-Fock (HF) 
wave function of the neutral atom ground state.  The exchange 
potential  ($V_{\rm exc}$) between the valence electrons and the HF 
core was computed without approximation.

The one-body and two-body polarization potentials ($V_{p1}$ and
$V_{p2}$) are semi-empirical with the short-range cut-off 
parameters derived by fitting to the spectra of their singly
ionized ions.  All details of the core-polarization potentials 
including the polarizabilities, $\alpha_d$, are given in 
\cite{bromley02a,bromley02b}.  Note that the functional form of
the polarization potential,$V_{p1}$, was set to be the same for 
the electrons and the positron. 

The positronic atom wave function is a linear combination 
of states created by multiplying atomic states to single 
particle positron states with the usual Clebsch-Gordan 
coupling coefficients ; 
\begin{eqnarray}
|\Psi;LS \rangle  &=& \sum_{i,j} c_{i,j} \ 
\langle L_i M_i \ell_j m_j|L M_L \rangle  
\langle S_i M_{S_i}  {\scriptstyle \frac{1}{2}} \mu_j|S M_S \rangle \nonumber \\  
&\times& \Phi_i(Atom;L_iS_i) \phi_j({\bf r}_0) \ .   
\end{eqnarray}
In this expression $\Phi_i(Atom;L_i S_i)$ is an antisymmetric 
atomic wave function with good $L$ and $S$ quantum numbers.  
The function $\phi_j({\bf r}_0)$ is a single positron orbital.
The single particle orbitals are written as a product of a 
radial function and a spherical harmonic:
\begin{equation}
\phi({\bf r})  =  P(r) Y_{lm}({\hat {\bf r}}) \ .         
\end{equation}
As the calculations were conducted in a fixed core model 
we used HF calculations of the neutral atom ground states
to construct the core orbitals.   These HF orbitals 
were computed with a program that can represent the radial 
wave functions as a linear combination of Slater Type
Orbitals (STO) \cite{mitroy99f}.  

A linear combination of STOs and Laguerre Types Orbitals (LTOs) 
was used to describe the radial dependence of electrons occupying
orbitals with the same angular momentum as those in the ground 
state.  Orbitals that did not have any core orbitals with
the same angular momentum were represented by a LTO set with
a common exponential parameter.     
The STOs give a good representation of the wave function 
in the interior region while the LTOs largely span the valence 
region.  The LTO basis \cite{bromley02a,bromley02b} has the 
property that the basis can be expanded toward completeness 
without introducing any linear independence problems.  
 
The CI basis included all the possible $L = 0$ configurations 
that could be formed by letting the two electrons and positron 
populate the single particle orbitals subject to two selection 
rules,
\begin{eqnarray}
\max(\ell_0,\ell_1,\ell_2) & \le & L_{\rm max} \ ,  \\ 
\min(\ell_1,\ell_2) & \le & L_{\rm int} \ .   
\end{eqnarray}
In these rules $\ell_0$ is the positron orbital angular momentum, 
while $\ell_1$ and $\ell_2$ are the angular momenta of the 
electrons.  A large value of $L_{\rm max}$ is necessary as the
attractive electron-positron interaction causes a pileup of 
electron density in the vicinity of the positron.  The $L_{\rm int}$ 
parameter was used to eliminate configurations involving the 
simultaneous excitation of both electrons into high $\ell$ states. 
Calculations on PsH and $e^+$Be  had shown that the choice 
of $L_{\rm int} = 3$ could reduce the dimension of the CI basis 
by a factor of 2 while having an effect of about 1$\%$ upon 
the binding energy and annihilation rate \cite{bromley02a}. 
The present set of calculations were all performed with
$L_{\rm int} = 4$.

Various expectation values were computed to provide information 
about the structure of these systems.  The mean distance of 
the electron and positron from the nucleus are denoted by
$\langle r_e \rangle$ and $\langle r_p \rangle$.  
The $2\gamma$ annihilation rate for annihilation with the 
core and valence electrons was computed with the usual 
expressions \cite{neamtan62,drachman95b,ryzhikh99a}.
The $2\gamma$ rate for the core ($\Gamma_c$) and 
valence ($\Gamma_v$) electrons are tabulated separately.  

\subsection{Extrapolation issues}

The feature that differentiates mixed electron-positron CI calculations 
from purely electron CI calculations is the slow convergence of the 
calculation with respect to $L_{\rm max}$, the maximum $\ell$ of any 
electron or positron orbital included in the CI basis.  Typically, a 
calculation is made to $L_{\rm max} \approx 10$ (or greater), with various 
extrapolation techniques used to estimate the $L_{\rm max} \to \infty$ 
correction.  For any expectation value one can write formally  
\begin{equation}
\langle X \rangle^{L_{\rm max}} = \sum_{L=0}^{L_{\rm max}} \Delta X^{L} \ ,
\label{XJ1}
\end{equation}
where $\Delta X^{L}$ is the increment to the observable that occurs
when the maximum orbital angular momentum is increased from
$L\!- \! 1$ to $L$, e.g.
\begin{equation}
\Delta X^{L} = \langle X \rangle^{L} \ - \ \langle X \rangle^{L-1} \ .
\label{XJ3}
\end{equation}
Hence, one can write formally
\begin{equation}
\langle X \rangle = \langle X \rangle^{L_{\rm max}}  \ + \sum_{L=L_{\rm max}+1}^{\infty} \Delta X^{L} \ .
\label{XJ2}
\end{equation}

However, it is quite easy to make substantial errors in estimating 
the $L_{\rm max} \to \infty$ correction \cite{mitroy05i,mitroy06a,mitroy06b}. 
There have been a number of investigations of the convergence of CI expansions
for electronic and mixed electron-positron systems  
\cite{schwartz62a,carroll79a,hill85a,kutzelnigg92a,schmidt93a,ottschofski97a,gribakin02a,mitroy02b,bromley06a,mitroy06a}.   
The reliability of the different methods to estimate the $L_{\rm max} \to \infty$ 
correction for the energy and annihilation rate has been assessed 
in detail elsewhere \cite{mitroy06a}.   In this work, only the briefest 
description of the recommended methods are described.    

The recent computational investigations of helium \cite{bromley06a} and
some positron-atom systems \cite{mitroy06a} suggest that usage of an
inverse power series of the generic type 
\begin{eqnarray}
\Delta X^{L_{\rm max}} &=& \frac{B_X}{(L_{\rm max}+{\scriptstyle \frac{1}{2}})^n}
   + \frac{C_X}{(L_{\rm max}+{\scriptstyle \frac{1}{2}})^{n+1}} \nonumber \\ 
   & + & \frac{D_X}{(L_{\rm max}+{\scriptstyle \frac{1}{2}})^{n+2}} + \ldots \ ,  
\label{Aseries} 
\end{eqnarray}
is the best way to determine the $L_{\rm max} \to \infty$ correction for 
the energy, $E$ and the 2$\gamma$ annihilation rate.  A three term 
series with $n = 4$ is used for the energy.   One needs four successive values 
of $E^{L}$ to determine the coefficients $B_E$, $C_E$ and $D_E$.
Once the coefficients have been fixed, the inverse power series is summed to 
$J_{\rm max} = 100$,after which the approximate result 
\begin{equation}
\sum_{L=J_{\rm max}+1}^{\infty} \frac{1}{(L+{\scriptstyle \frac{1}{2}})^p } 
          \approx  \frac{1}{(p-1)(J_{\rm max}+1)^{p-1} }  \ , 
\label{bettertail}
\end{equation}
is used \cite{mitroy06b}.  

The correction to $\Gamma$ follows the same general procedure as 
the energy, but with two differences.  The power in eq.~(\ref{Aseries})   
is set to $n = 2$ and only 2-terms are retained in the series   
(requiring three successive values of $\Gamma^{L}$). 

The usage of the inverse power series is the preferred approach 
when the asymptotic form for $\Delta X^L$ has been established 
by perturbation theory.  For other operators it is best to    
to use a single-term inverse power series with an indeterminate
power, e.g
\begin{equation}
\Delta X^L  = \frac{A}{(L+ {\scriptstyle \frac{1}{2}})^p} .  
\end{equation}
The factors $A$ and $p$ can be determined from the three largest calculations 
using 
\begin{equation}
p =   \ln \left(  \frac {\Delta X^{L_{\rm max}-1}}{\Delta X^{L_{\rm max}}} \right) \biggl/
      \ln \left( \frac{L_{\rm max}+{\scriptstyle \frac{1}{2}}}{L_{\rm max}-{\scriptstyle \frac{1}{2}}} \right) \ , 
\label{pdef}
\end{equation}
and 
\begin{equation}
A =   \Delta X^{L_{\rm max}} (L_{\rm max} +{\scriptstyle \frac{1}{2}})^{p}  \ .
\label{Avalue}
\end{equation}
Once $p$ and $A$ are determined,  
the $L_{\rm max} \to \infty$ correction can be included using the
same procedure as adopted for the multi-term fits to the energy and
annihilation.  
This method is used in determination of the $L_{\rm max} \to \infty$ estimates 
of $\langle r_e \rangle$,  $\langle r_p \rangle$ and $\Gamma_c$.  However, the 
value of $p$ is computed for all operators since it is useful to know whether 
$p_E$ and $p_{\Gamma_v}$ are close to the expected values of 4 and 2 respectively.  
While the subdivision of the annihilation rate into core and valence 
components is convenient for physical interpretation, it was also done on
mathematical grounds.  The calculation of $\Gamma_c$ does not explicitly 
include correlations between the core electrons and the positron, and so
the $\Delta \Gamma_c^L$ increments converge faster than   
the $\Delta \Gamma_v^L$ increments (i.e. $p_{\Gamma_c} > p_{\Gamma_v}$). 

\section{Calculation Results}

\subsection{Improved FCSVM data for $e^+$Be and $e^+$Mg}

The FCSVM \cite{ryzhikh98e,mitroy02b} 
has also been applied to determine the structures of  
$e^+$Be and $e^+$Mg \cite{ryzhikh98e,mitroy01c}.  The FCSVM
expands the wave function as a linear combination of 
explicitly correlated gaussians (ECGs), with the core orbitals
taken from a HF calculation.  One- and two-body polarization
potentials are included while orthogonality of
the active electrons with the core is enforced by the use of an 
orthogonalizing pseudo-potential \cite{ryzhikh98e,mitroy99h,mitroy02b}.
The FCSVM model hamiltonians are very similar to those used in
the CI calculations.  But there are some small differences in
detail that lead to the FCSVM hamiltonian giving slightly
different energies.      

The best previous FCSVM wave function for $e^+$Be \cite{mitroy01c} gave 
a binding energy, 0.03147 hartree, and annihilation rate 
$0.420 \times 10^9$ sec$^{-1}$,that were close to convergence.  
Some extensive re-optimizations seeking to improve the quality of 
the wave function in the asymptotic region yielded only minor changes 
(of the order of 1$\%$) in the ground state properties \cite{mitroy05d}.  
Nevertheless, the latest energies and expectation values for the 
$e^+$Be ground state are tabulated in Tables \ref{Beptab} and 
\ref{summary}.  These values should be converged to better
than 1$\%$ with respect to further enlargement and optimization
of the ECG basis.    

The more complex core for Mg does slow the convergence of the energy and 
other properties of $e^+$Mg considerably \cite{mitroy02b}.  The best 
energy previously reported for this system was 0.016096 hartree 
\cite{mitroy05d}.  The current best wave function, which is constructed
from a linear combination of 1200 ECGs gives a binding energy of 0.016930 
hartree and a valence annihilation rate of $1.0137 \times 10^9$ sec$^{-1}$.  
Other expectation values are listed in Table \ref{Beptab}.  Examination
of the convergence pattern during the series of basis set enlargements
and optimizations suggests that the binding energy and annihilation rate 
are converged to between 2$\%$ and 5$\%$.  
 
The FCSVM binding energies do have a weak dependence on one parameter 
in the calculation since the orthogonalizing pseudo-potential is  
actually a penalty function, viz  
\begin{equation}      
\lambda {\hat P} = \sum_{i \ \in \ \text{core}} \lambda |\phi_i \rangle \langle \phi_i | \ ,  
\label{OPP} 
\end{equation}      
that was added to the hamiltonian.  Choosing $\lambda$ to be large and positive 
means the energy minimization automatically acts to construct a wave function 
which has very small overlap with the core \cite{krasnopolsky74,ryzhikh98e,mitroy99h}.  
The FCSVM properties reported in Tables \ref{Beptab} and \ref{summary} were 
computed with $\lambda = 10^5$ hartree.  The core overlap (i.e. the expectation 
value of ${\hat P}$) was $1.86 \times 10^{-11}$ for $e^+$Be and 
$1.61 \times 10^{-10}$ for $e^+$Mg.    

\subsection{CI results for group II atoms}

Table \ref{Beptab} contains the results of the current series of
calculations on the four positronic atoms.  
The size of the calculations for the four atoms were almost 
the same. The electron-electron angular momentum selector was set to 
$L_{\rm int} = 4$.  For $\ell > 3$ at least 15 LTOs were included in 
the radial basis sets for the electron and positron orbitals.  For 
$\ell \le 2$ the dimension of the orbital basis sets were slightly
larger than 15 and the basis sets for electrons occupying orbitals
with the same angular momentum as those in the core were typically a 
mix of STOs (to describe the electron close to nucleus) and LTOs. The 
calculations used basis sets with $L_{\rm max} = 9, 10, 11$ and 12.  
The calculations with $L_{\rm max} < 12$ had configuration spaces 
which were subsets of the $L_{\rm max} = 12$ and this expedited 
the computations since one list of radial matrix elements was 
initially generated for the $L_{\rm max} = 12$ basis and then reused 
for the smaller basis sets.    
  
The secular equations that arose typically had dimensions of about 
500,000 and the diagonalizations were performed with the Davidson 
algorithm using a modified version of the program of Stathopolous 
and Froese-Fischer \cite{stathopolous94a}.  Convergence was not very 
quick and about 16000 iterations were needed to achieve convergence 
in some cases.  It was possible to speed up the diagonalization for 
$L_{\rm max} < 12$.  An edited eigenvector from the $L_{\rm max} = 12$ 
calculation was used as the initial eigenvector estimate, and this 
often reduced the number of iterations required by 50$\%$.
  
\subsubsection{Results for $e^+$Be}

The lowest energy dissociation channel is the $e^+ + \text{Be}$ 
channel, which has an energy of $-1.01181167$ hartree with respect to 
the doubly ionized Be$^{2+}$ core.   The agreement of the 
extrapolated CI binding energy of $\varepsilon = 0.003169$  
hartree with the FCSVM binding energy of $\varepsilon  =0.003180$ 
is better than 1$\%$.   A similar level of agreement exists for 
the $\langle r_e \rangle$ and $\langle r_p \rangle$ expectation
values. 

The only expectation value for which 1$\%$ level of agreement 
does not occur is the
annihilation rate and here the extrapolated CI value of 
$0.4110 \times 10^9$ sec$^{-1}$ is only about $3.5\%$ smaller 
than the FCSVM of $0.4267 \times 10^9$ sec$^{-1}$.  However,
it is known that the convergence of the annihilation rate
with respect to an increasing number of radial basis functions
is slower than the convergence of the energy 
\cite{mitroy06a,bromley06a}.  This means that a CI type calculation
has an inherent tendency to underestimate the annihilation rate.    
For example, a CI calculation on PsH of similar size to the 
present $e^+$Be calculation underestimated the annihilation
rate by 6$\%$ \cite{mitroy06a}.  That the exponent of the polar law 
decay, $p_{\Gamma_v} = 2.10$, is larger than the expected asymptotic 
value of $p = 2.0$ is consistent with this idea.   A better estimate 
of the annihilation rate can be obtained by simply forcing
$C_{\Gamma}$ to be zero in eq.~(\ref{Aseries}) and thus using 
$\Delta \Gamma^{12}$ to fit $B_{\Gamma}$.  When this is done 
done the annihilation rate increases to $0.4178 \times 10^9$ 
sec$^{-1}$.
    
\begin{table*}[th]
\caption[]{  Results of CI calculations for positronic alkaline-earth 
atoms for a given $L_{\rm max}$.  The $E$ column gives the three-body energy 
with respect to the doubly ionized frozen core and $\varepsilon$ is the 
binding energy with respect to the lowest energy dissociation channel.  
The $\Gamma_v$ and $\Gamma_c$ columns give the valence and core 
annihilation rate (in $10^9$ sec$^{-1}$) .  The results in the row $10^*$ 
are taken from earlier CI calculations for these systems 
\cite{bromley02a,bromley02b} with $L_{\rm max} = 10$ (and 
$L_{\rm int} = 3$).  The results in the row $\infty$ use the methods 
described in the body of the text to evaluate the 
$L_{\rm max} \rightarrow \infty$ correction.  The exponent $p$ 
characterizes the rate of decay of the expectation value increments 
evaluated at $L_{\rm max} = 12$ using eq.~(\ref{pdef}). 
}
\label{Beptab}
\vspace{0.5cm}
\begin{ruledtabular}
\begin{tabular}{lcccccc}
$L_{\rm max}$&  $E$ &  $\varepsilon$
                   &  $\langle r_e \rangle$ &  $\langle r_p \rangle$ & $\Gamma_c$  & $\Gamma_v$  \\ \hline
\multicolumn{7}{c}{$e^+$Be}  \\
10$^*$   & (-1.0143769)& (0.002533) & (2.639)   & (10.746)  & (0.001962) & (0.2411)   \\
9        & -1.01435756 & 0.00254589 & 2.6388477 & 10.874256 & 0.00193993 & 0.24026720 \\
10       & -1.01448318 & 0.00267151 & 2.6418168 & 10.699433 & 0.00198405 & 0.25651443 \\
11       & -1.01457837 & 0.00276670 & 2.6441227 & 10.574208 & 0.00201619 & 0.27004634 \\
12       & -1.01465138 & 0.00283971 & 2.6459282 & 10.482126 & 0.00204005 & 0.28140404 \\
$p$      & 3.1806     & 3.1806    & 2.9339   & 3.6871   & 3.5764     & 2.1006    \\
$\infty$ & -1.0149809  & 0.0031692 & 2.65673 & 10.09755  & 0.002144 & 0.410976  \\
FCSVM    & -1.0151335  &  0.003180  & 2.654     & 10.048    & 0.00221    & 0.4267 \\
\hline 
\multicolumn{7}{c}{$e^+$Mg}  \\
10$^*$   & (-0.8473592)& (0.0145092)& (3.382)   & (7.101)  & (0.010845) & (0.5429) \\ 
9        & -0.84741494 & 0.01450067 & 3.3831320 & 7.116532 & 0.01079647 & 0.54089010 \\
10       & -0.84790548 & 0.01499121 & 3.3936654 & 7.071950 & 0.01084944 & 0.57692369 \\
11       & -0.84828090 & 0.01536663 & 3.4022694 & 7.040929 & 0.01087921 & 0.60775407 \\
12       & -0.84857204 & 0.01565777 & 3.4093312 & 7.018703 & 0.01089568 & 0.63435278 \\
$p$      & 3.0496     &  3.0496   & 2.3690    & 3.9985  & 7.0953     & 1.7706 \\
$\infty$ & -0.8499543  & 0.0170400 & 3.47039 & 6.93657   & 0.010922 & 0.990069 \\
FCSVM    & -0.849002   &  0.016930  & 3.447     & 6.923    & 0.0112     & 1.0137 \\
\hline 
\multicolumn{7}{c}{$e^+$Ca}  \\
10$^*$   & (-0.6986443)& (0.0123578)& (4.456)   & (6.848)  & (0.01355)  & (0.7335)   \\
9        & -0.69855551 & 0.01226895 & 4.4602428 & 6.863740 & 0.01343426 & 0.72709017 \\
10       & -0.69975764 & 0.01347109 & 4.4873848 & 6.872414 & 0.01323075 & 0.78001274 \\
11       & -0.70069553 & 0.01440898 & 4.5110869 & 6.885039 & 0.01304512 & 0.82640757 \\
12       & -0.70143637 & 0.01514981 & 4.5315631 & 6.898804 & 0.01288316 & 0.86733542 \\
$p$      & 2.8286     &   2.8286  & 1.7546    & -1.0371 & 1.6361    & 1.5037  \\
$\infty$ & -0.7052160 & 0.0189295  & 4.86076   & ---      & 0.009780   & 1.478148   \\
\hline   
\multicolumn{7}{c}{$e^+$Sr}  \\
10$^*$   & (-0.6602186)& (0.0048689)& (4.850)   & (7.056)  & (0.01487)  & (0.7488) \\
9        & -0.65997599 & 0.00462598 & 4.8638673 & 7.100141 & 0.01464684 & 0.73239378 \\
10       & -0.66146709 & 0.00611708 & 4.8979559 & 7.123685 & 0.01432317 & 0.78845209 \\
11       & -0.66263875 & 0.00728874 & 4.9283728 & 7.150071 & 0.01403253 & 0.83790890 \\
12       & -0.66357065 & 0.00822064 & 4.9552753 & 7.176708 & 0.01377785 & 0.88177286 \\
$p$      & 2.7459      & 2.7459     & 1.4725   & -0.1134  & 1.5844    & 1.4393  \\
$\infty$ & -0.6684520  & 0.0131020  & 5.65380   & ---      & 0.008456   & 1.552589   \\
\end{tabular}
\end{ruledtabular}
\end{table*}

\subsubsection{Results for $e^+$Mg}

The results of the calculations with $e^+$Mg are listed in Table 
\ref{Beptab}.  The lowest energy dissociation channel is 
to $e^+ + \text{Mg}$, which has an energy of $-0.83291427$ hartree
with respect to the doubly ionized Mg$^{2+}$ core.   

The CI calculations, reported in Table \ref{Beptab} for 
$L_{\rm max} =$ 9, 10, 11 and 12 are largely consistent with 
the FCSVM calculations.  The largest explicit CI calculation gives a 
binding energy of 0.015658 hartree.  Extrapolation to the
$L_{\rm max} \rightarrow \infty$ limit adds about $10\%$ to 
the binding energy, and the final estimate was 0.017040 hartree. 
Despite the better than 1$\%$ agreement between the CI and
FCSVM calculations, a further binding energy increase of about 
1-2$\%$ would be conceivable if both calculations were taken 
to the variational limit.        

The slow convergence of $\Gamma_v$ with $L_{\rm max}$ is evident
from Table \ref{Beptab} and the extrapolation correction 
contributes about 36$\%$ to the overall annihilation rate.  The
present $L_{\rm max} \to \infty$ estimate can be expected to be 
too small by 5-10$\%$. 

All the other expectation values listed in Table \ref{Beptab} lie
with 1-2$\%$ of those of the FCSVM expectation values.  As a
general rule, inclusion of the $L_{\max} \to \infty$ corrections 
generally improves the agreement between the CI and FCSVM 
calculations.  
   
\subsubsection{Results for $e^+$Ca}

The results of the calculations with $e^+$Ca are listed in Table 
\ref{Beptab}.  Since neutral calcium has an ionization potential 
smaller than the energy of Ps ground state (the present model 
potential and electron orbital basis gives -0.43628656 hartree for 
the Ca$^+$ energy and -0.65966723 hartree for the neutral Ca energy),  
its lowest energy dissociation channel is the $\text{Ps} + \text{Ca}^+$ 
channel.  The present model potential gives this channel an energy of 
$-0.68628656$ hartree.  

The energies listed in Table \ref{Beptab} indicate that $e^+$Ca is 
the positronic atom with the largest known binding energy, namely  
$\varepsilon = 0.018929$ hartree.  The $L_{\rm max} \to \infty$ 
correction contributes 20 $\%$ of the binding energy.  The partial 
wave series is more slowly convergent for $e^+$Ca than for $e^+$Mg
(i.e. $p_E$ is smaller, and the coefficients $C_E$ and $D_E$ in eq.~(\ref{bettertail}) 
are larger).  This is expected since calcium has a smaller ionization 
potential, and so the electrons are located a greater distance away 
from the nucleus.  This makes it easier for the positron to attract the 
electrons, and the stronger pileup of electron density around the positron 
further from the nucleus requires a longer partial wave expansion 
to represent correctly.   

The slower convergence of the wave function with $L_{\rm max}$ makes an even 
larger impact on the annihilation rate.  Some 41$\%$ of the annihilation rate 
of $\Gamma_v = 1.478 \times 10^9$ sec$^{-1}$ comes from the $L_{\rm \max} \to \infty$ 
correction.  As mentioned earlier for $e^+$Mg, it is likely that this value is slightly
smaller than the true annihilation rate. 

The extrapolation corrections for $\langle r_p \rangle$ and 
$\Gamma_c$ listed in Table \ref{Beptab} are unreliable.  
The $e^+$Ca system, at large distances consists of Ca$^+$ + Ps. In 
other calculations of positron binding systems it has been 
noticed that systems that decay asymptotically into
$\text{Ps} + \text{X}$ do not have an $\langle r_p \rangle$ that 
changes monotonically with $L_{\rm max}$ \cite{bromley00a,bromley02a}. 
Initially, the positron becomes more tightly bound to the system 
as $L_{\rm max}$ increases, resulting in a decrease in 
$\langle r_p \rangle$.  However, $\langle r_p \rangle$ tends to 
increase at the largest values of $L_{\rm max}$.   The net result of 
all this is that $\Delta \langle r_p \rangle^{L}$ (and by 
implication $\Delta \Gamma_c^L$) approach their asymptotic forms very 
slowly. The best policy is to simply not to give any credence to 
the extrapolation corrections for either of these operators for 
$e^+$Ca (and $e^+$Sr).  The small value of $p$ for 
$\Delta \langle r_e \rangle^{L}$ suggests that the reliability of the  
$L_{\rm max} \to \infty$ correction may be degraded for this
expectation value as well.    
   
\subsubsection{Results for $e^+$Sr}

The results of the calculations for  $e^+$Sr are listed in Table 
\ref{Beptab}.  Since neutral strontium has an ionization potential 
smaller than the energy of Ps ground state (the present model 
potential and electron orbital basis gives -0.40535001 hartree for 
the Sr$^+$ energy and -0.61299101  hartree for the neutral Sr energy),  
its lowest energy dissociation channel is the $\text{Ps} + \text{Sr}^+$ 
channel, which has an energy of -0.65535001 hartree.  The small 
ionization potential of 0.20764100 hartree means that the structure 
of the $e^+$Sr ground state will be dominated by a 
$\Psi(\text{Sr}^+$)$\Psi(\text{Ps})$ type configuration \cite{mitroy02b}.  
This leads to slower convergence of the ground state with 
$L_{\rm max}$ which is evident from Table \ref{Beptab}. 

As expected, the binding energy of $e^+$Sr is smaller than that of 
$e^+$Ca.  Previous investigations have indicated that positron binding 
energies should be largest for atoms with ionization potentials closest 
to 0.250 hartree (the Ps binding energy) \cite{mitroy99b,mitroy02f}.  
There is obviously some uncertainty in the precise determination of the 
binding energy due to fact that $L_{\rm max} \to \infty$ correction 
constitutes some $37 \%$ of the binding energy of 0.013102 hartree.  
The net effect of errors due to the extrapolation correction are not 
expected to be excessive.  Applying eq.~(\ref{Aseries}) with only the 
first two-terms retained (i.e. $D_E = 0$) results in a final energy 
0.012764 hartree, which is 3$\%$ smaller than the value of 0.013102
Hartree.   The present $e^+$Sr binding energy is some 30$\%$ larger
than the energy of the previous CI calculation listed in Table 
\ref{summary} \cite{bromley02b}.  

The final estimate of the valence annihilation rate was 
$ 1.553 \times 10^9$ sec$^{-1}$ and some 43$\%$ of the annihilation rate 
comes from the $L_{\rm \max} \to \infty$ correction.  This value of 
$\Gamma_v$ could easily be 10$\%$ smaller than the true annihilation rate.  
The explicitly calculated expectation values for $\langle r_e \rangle$,   
$\langle r_p \rangle$ and $\Gamma_c$ at $L_{\rm max} =12$ should be 
preferred since the $L_{\rm \max} \to \infty$ corrections in these 
cases are likely to be unreliable. 

\subsection{3-body clustering}

While the truncation of the basis to $L_{\rm int} = 4$ has little
effect on the $e^+$Be system, its effect is larger for the $e^+$Sr 
system.   The more loosely bound alkaline-earth atoms have their
electrons localized further away from the nucleus, and this makes it
easier for the positron to form something like a Ps$^{-}$ cluster 
\cite{mitroy02f,mitroy04a}.   When this occurs, correlations of the 
positron with {\em both} electrons increase in strength, and the 
inclusion of configurations with $L_{\rm int} > 4$ becomes more 
important.  

The relative size of of these neglected $L_{\rm int} > 4$ configurations 
can be estimated using techniques similar to those adopted for the   
$L_{\rm max} \to \infty$ corrections.  Calculations for a succession
of $L_{\rm int}$ values were performed in earlier works 
\cite{bromley02a,bromley02b}. The assumption is made that the
binding energy and annihilation rate increments scale as 
$A/(L_{\rm int}+{\scriptstyle \frac{1}{2}})^4$ (note, the 
power of 4 for the annihilation is used since $L_{\rm int}$ only 
has a directly effect on electron-electron correlations).  The 
difference between an $L_{\rm int} = 2$ and $L_{\rm int} = 3$ 
is used to estimate $A$ and then eq.(~\ref{bettertail}) 
determines the $L_{\rm int} \to \infty$ correction
(in the case of $e^+$Be calculations up to $L_{\rm int} = 10$ 
exist \cite{bromley02a}).     

Table \ref{summary} contains a summary of the final binding energies 
obtained from the present CI calculations, and earlier binding energies 
obtained alternate methods.  As part of this table, energies with an 
additional $L_{\rm int} \to \infty$ correction are also given.  The size 
of the correction ranges from $1.8 \times 10^{-5}$ hartree for $e^+$Be 
to $21.9 \times 10^{-5}$ hartree for $e^+$Sr.  Even though these 
estimations of the correction are not rigorous, they indicate that  
the underestimation in the binding energy resulting from a truncation 
of the configuration space to $L_{\rm int} < 4$ is most likely to be 2$\%$ 
or smaller.    

A similar analysis could be done for the annihilation rate but previous
results indicate that $\Gamma_v$ is less sensitive than $\varepsilon$ 
to an increase in $L_{\rm int}$ \cite{bromley02a,bromley02b}.  The net
increases in $\Gamma_v$ for $e^+$Be, $e^+$Mg, $e^+$Ca and $e^+$Sr were
$0.0011 \times 10^9$ sec$^{-1}$, $0.0030 \times 10^9$ sec$^{-1}$, 
$0.0039 \times 10^9$ sec$^{-1}$ and  $0.0039 \times 10^9$ sec$^{-1}$ 
respectively.  All of these extra contributions to $\Gamma_v$ correspond 
to changes of less than 0.5$\%$.

\begin{table}
\caption[] { \label{summary}
Binding energies (in hartree) of positronic beryllium, magnesium, 
calcium and strontium.  Only the latest calculations of a given type 
by a particular group are listed in this table.  
}
\vspace{0.2cm}
\begin{ruledtabular}
\begin{tabular}{lcccc}
Calculation  &   $e^+$Be & $^+e$Mg  & $e^+$Ca &   $e^+$Sr \\  \hline  
CI ($L_{\rm max}=12$) & 0.002840   &    0.015658   &  0.015150    &  0.008221    \\
CI  ($L_{\rm max} \to \infty$)  & 0.003169   &    0.017040   &  0.018929    &  0.013102    \\  
CI  ($L_{\rm int} \to \infty$)  & 0.003187   &    0.017099   &  0.019122    &  0.013321   \\  
Previous-CI \footnotemark[1] & 0.003083   &    0.01615   &  0.01650    &  0.01005   \\ 
FCSVM   & 0.003161   &    0.016930   &      &     \\  
DMC \footnotemark[2] & 0.0012(4)  &    0.0168(14)  &      &     \\
SVM \footnotemark[3]  & 0.001687   &           &      &     \\
PO \footnotemark[4]   &       &  0.00055   &    &     \\      
PO \footnotemark[5]   &       &  0.00459   &    &     \\      
MBPT \footnotemark[6]   &       &  0.0362    &    &     \\      
\end{tabular}
\end{ruledtabular}
\footnotetext[1]{Previous CI ($L_{\rm max}\to \infty$) \cite{bromley02a,bromley02b} } 
\footnotetext[2]{DMC, the statistical uncertainty in the last digit(s) is given in the brackets \cite{mella02a}}     
\footnotetext[3]{Fully ab-initio SVM \cite{ryzhikh98e} }  
\footnotetext[4]{Polarized orbital calculation, dipole only \cite{szmytkowski93a} }  
\footnotetext[5]{Polarized orbital calculation \cite{mceachran98} }  
\footnotetext[6]{Many Body perturbation theory  \cite{gribakin96} } 
\end{table}

\section{Summary and conclusions}

The summary of binding energies, produced by the current methods and 
other completely different approaches presented in Table \ref{summary} shows 
that the only methods that consistently agree with each other are the CI 
and FCSVM calculations.  Both these methods are variational in nature, 
both use realistic model potentials designed on very similar lines, and both 
have shown a tendency for the binding energies to slowly creep upwards 
as the calculation size is increased (refer to refs 
\cite{ryzhikh98e,mitroy02b,bromley02b} for examples of earlier and
slightly smaller binding energies).   The PO and MBPT approaches do not 
give reliable binding energies.      

The diffusion Monte Carlo method \cite{mella02a} gives an $e^+$Mg binding 
energy of 0.0168$\pm0.0014$ hartree which is very close to the present energy.  
This calculation was fully \textit{ab-initio} and did not use the fixed core 
approximation.  However, application of the same diffusion Monte Carlo
method to $e^+$Be gave a binding energy which is only half the size of the 
present value.  

The present binding energies are all larger than those given previously 
\cite{bromley02a,bromley02b} due to the usage of a radial basis which was 
almost twice the size of earlier calculations.  In two cases, $^+$Ca and 
$e^+$Sr the increase in binding energy exceeds 10$\%$.  The binding energies 
for $e^+$Be and $e^+$Mg are in agreement with those of FCSVM calculations to 
within their mutual uncertainties.  
Further enlargement of the basis could lead to the positron binding energies 
for Mg, Ca and Sr increasing by a few percent.           

Estimates of the annihilation rate have also been extracted from the CI
wave functions.  The present annihilation rates are certainly underestimates
of the true annihilation rate.   The annihilation rate converges very
slowly with respect to the radial basis and similar sized calculations 
on PsH suggest that the present annihilation rates could easily be too small
by at least 5$\%$ \cite{bromley06a,mitroy06a,mitroy06c}.   

The speed at which the partial wave expansion converges with respect to 
$L_{\rm max}$ is seen to decrease as the ionization energy of the parent 
atom decreases \cite{bromley02b,mitroy02b}.  In addition, the importance of 
3-body clustering (i.e. convergence with respect to $L_{\rm int}$) was
seen to increase as the ionization energy of the parent atom decreased 
\cite{mitroy02f}.     

The main factor limiting the size of the calculations now is 
the time taken to perform the diagonalizations.   Although, the 
calculations were performed on a Linux/Myrinet-based cluster,  
the sheer number of iterations, (16000 in the worst case), used by 
the Davidson method, meant that it could take 30 days to perform a  
diagonalization using 24 CPUs.  However, the main reason for 
adopting the Davidson method was the availability of a program 
that was easy to modify \cite{stathopolous94a}.  Usage of the 
more general Lanczos method \cite{whitehead77a} might lead to a 
quicker diagonalization and thus permit even larger calculations.    
  
\begin{acknowledgments}

This work was supported by a research grant from the Australian 
Research Council.  The calculations were performed on a Linux 
cluster hosted at the South Australian Partnership for Advanced 
Computing (SAPAC) with thanks to Grant Ward, Patrick Fitzhenry
and John Hedditch for their assistance.  The authors would like to 
thank Shane Caple for providing workstation maintenance and arranging 
access to additional computing resources. 
\end{acknowledgments}


\begin{thebibliography}{43}
\expandafter\ifx\csname natexlab\endcsname\relax\def\natexlab#1{#1}\fi
\expandafter\ifx\csname bibnamefont\endcsname\relax
  \def\bibnamefont#1{#1}\fi
\expandafter\ifx\csname bibfnamefont\endcsname\relax
  \def\bibfnamefont#1{#1}\fi
\expandafter\ifx\csname citenamefont\endcsname\relax
  \def\citenamefont#1{#1}\fi
\expandafter\ifx\csname url\endcsname\relax
  \def\url#1{\texttt{#1}}\fi
\expandafter\ifx\csname urlprefix\endcsname\relax\def\urlprefix{URL }\fi
\providecommand{\bibinfo}[2]{#2}
\providecommand{\eprint}[2][]{\url{#2}}

\bibitem[{\citenamefont{Mitroy et~al.}(2002)\citenamefont{Mitroy, Bromley, and
  Ryzhikh}}]{mitroy02b}
\bibinfo{author}{\bibfnamefont{J.}~\bibnamefont{Mitroy}},
  \bibinfo{author}{\bibfnamefont{M.~W.~J.} \bibnamefont{Bromley}},
  \bibnamefont{and} \bibinfo{author}{\bibfnamefont{G.~G.}
  \bibnamefont{Ryzhikh}}, \bibinfo{journal}{J.~Phys.~B}
  \textbf{\bibinfo{volume}{35}}, \bibinfo{pages}{R81} (\bibinfo{year}{2002}).

\bibitem[{\citenamefont{Schrader}(2001)}]{schrader01a}
\bibinfo{author}{\bibfnamefont{D.~M.} \bibnamefont{Schrader}}, in
  \emph{\bibinfo{booktitle}{New Directions in Antimatter Physics and
  Chemistry}}, edited by \bibinfo{editor}{\bibfnamefont{C.~M.}
  \bibnamefont{Surko}} \bibnamefont{and} \bibinfo{editor}{\bibfnamefont{F.~A.}
  \bibnamefont{Gianturco}} (\bibinfo{publisher}{Kluwer Academic Publishers},
  \bibinfo{address}{The Netherlands}, \bibinfo{year}{2001}), p.
  \bibinfo{pages}{263}.

\bibitem[{\citenamefont{Strasburger and Chojnacki}(2003)}]{strasburger03a}
\bibinfo{author}{\bibfnamefont{K.}~\bibnamefont{Strasburger}} \bibnamefont{and}
  \bibinfo{author}{\bibfnamefont{H.}~\bibnamefont{Chojnacki}}, in
  \emph{\bibinfo{booktitle}{Explicitly Correlated Wave Functions in Chemistry
  and Physics: Theory and Applications}}, edited by
  \bibinfo{editor}{\bibfnamefont{J.}~\bibnamefont{Rychlewski}}
  (\bibinfo{publisher}{Kluwer Academic Publishers}, \bibinfo{address}{The
  Netherlands}, \bibinfo{year}{2003}), p. \bibinfo{pages}{439}.

\bibitem[{\citenamefont{Mitroy}(2002)}]{mitroy02f}
\bibinfo{author}{\bibfnamefont{J.}~\bibnamefont{Mitroy}},
  \bibinfo{journal}{Phys.~Rev.~A} \textbf{\bibinfo{volume}{66}},
  \bibinfo{pages}{010501} (\bibinfo{year}{2002}).

\bibitem[{\citenamefont{Ryzhikh and Mitroy}(1998)}]{ryzhikh98c}
\bibinfo{author}{\bibfnamefont{G.~G.} \bibnamefont{Ryzhikh}} \bibnamefont{and}
  \bibinfo{author}{\bibfnamefont{J.}~\bibnamefont{Mitroy}},
  \bibinfo{journal}{J.~Phys.~B} \textbf{\bibinfo{volume}{31}},
  \bibinfo{pages}{L401} (\bibinfo{year}{1998}).

\bibitem[{\citenamefont{Ryzhikh
  et~al.}(1998{\natexlab{a}})\citenamefont{Ryzhikh, Mitroy, and
  Varga}}]{ryzhikh98e}
\bibinfo{author}{\bibfnamefont{G.~G.} \bibnamefont{Ryzhikh}},
  \bibinfo{author}{\bibfnamefont{J.}~\bibnamefont{Mitroy}}, \bibnamefont{and}
  \bibinfo{author}{\bibfnamefont{K.}~\bibnamefont{Varga}},
  \bibinfo{journal}{J.~Phys.~B} \textbf{\bibinfo{volume}{31}},
  \bibinfo{pages}{3965} (\bibinfo{year}{1998}{\natexlab{a}}).

\bibitem[{\citenamefont{Mitroy and Ryzhikh}(2001)}]{mitroy01c}
\bibinfo{author}{\bibfnamefont{J.}~\bibnamefont{Mitroy}} \bibnamefont{and}
  \bibinfo{author}{\bibfnamefont{G.~G.} \bibnamefont{Ryzhikh}},
  \bibinfo{journal}{J.~Phys.~B} \textbf{\bibinfo{volume}{34}},
  \bibinfo{pages}{2001} (\bibinfo{year}{2001}).

\bibitem[{\citenamefont{Ryzhikh
  et~al.}(1998{\natexlab{b}})\citenamefont{Ryzhikh, Mitroy, and
  Varga}}]{ryzhikh98b}
\bibinfo{author}{\bibfnamefont{G.~G.} \bibnamefont{Ryzhikh}},
  \bibinfo{author}{\bibfnamefont{J.}~\bibnamefont{Mitroy}}, \bibnamefont{and}
  \bibinfo{author}{\bibfnamefont{K.}~\bibnamefont{Varga}},
  \bibinfo{journal}{J.~Phys.~B} \textbf{\bibinfo{volume}{31}},
  \bibinfo{pages}{L265} (\bibinfo{year}{1998}{\natexlab{b}}).

\bibitem[{\citenamefont{Bromley and Mitroy}(2002{\natexlab{a}})}]{bromley02a}
\bibinfo{author}{\bibfnamefont{M.~W.~J.} \bibnamefont{Bromley}}
  \bibnamefont{and} \bibinfo{author}{\bibfnamefont{J.}~\bibnamefont{Mitroy}},
  \bibinfo{journal}{Phys.~Rev.~A} \textbf{\bibinfo{volume}{65}},
  \bibinfo{pages}{012505} (\bibinfo{year}{2002}{\natexlab{a}}).

\bibitem[{\citenamefont{Bromley and Mitroy}(2002{\natexlab{b}})}]{bromley02b}
\bibinfo{author}{\bibfnamefont{M.~W.~J.} \bibnamefont{Bromley}}
  \bibnamefont{and} \bibinfo{author}{\bibfnamefont{J.}~\bibnamefont{Mitroy}},
  \bibinfo{journal}{Phys.~Rev.~A} \textbf{\bibinfo{volume}{65}},
  \bibinfo{pages}{062505} (\bibinfo{year}{2002}{\natexlab{b}}).

\bibitem[{\citenamefont{Dzuba et~al.}(1999)\citenamefont{Dzuba, Flambaum,
  Gribakin, and Harabati}}]{dzuba99}
\bibinfo{author}{\bibfnamefont{V.~A.} \bibnamefont{Dzuba}},
  \bibinfo{author}{\bibfnamefont{V.~V.} \bibnamefont{Flambaum}},
  \bibinfo{author}{\bibfnamefont{G.~F.} \bibnamefont{Gribakin}},
  \bibnamefont{and} \bibinfo{author}{\bibfnamefont{C.}~\bibnamefont{Harabati}},
  \bibinfo{journal}{Phys.~Rev.~A} \textbf{\bibinfo{volume}{60}},
  \bibinfo{pages}{3641} (\bibinfo{year}{1999}).

\bibitem[{\citenamefont{Saito}(2003)}]{saito03a}
\bibinfo{author}{\bibfnamefont{S.~L.} \bibnamefont{Saito}},
  \bibinfo{journal}{J.~Chem.~Phys.} \textbf{\bibinfo{volume}{118}},
  \bibinfo{pages}{1714} (\bibinfo{year}{2003}).

\bibitem[{\citenamefont{Strasburger and Chojnacki}(1995)}]{strasburger95}
\bibinfo{author}{\bibfnamefont{K.}~\bibnamefont{Strasburger}} \bibnamefont{and}
  \bibinfo{author}{\bibfnamefont{H.}~\bibnamefont{Chojnacki}},
  \bibinfo{journal}{Chem.~Phys.~Lett.} \textbf{\bibinfo{volume}{241}},
  \bibinfo{pages}{485} (\bibinfo{year}{1995}).

\bibitem[{\citenamefont{Schrader}(1998)}]{schrader98}
\bibinfo{author}{\bibfnamefont{D.~M.} \bibnamefont{Schrader}},
  \bibinfo{journal}{Nucl.~Instrum.~Methods~Phys.~Res.~B}
  \textbf{\bibinfo{volume}{143}}, \bibinfo{pages}{209} (\bibinfo{year}{1998}).

\bibitem[{\citenamefont{Mitroy and Ryzhikh}(1999{\natexlab{a}})}]{mitroy99c}
\bibinfo{author}{\bibfnamefont{J.}~\bibnamefont{Mitroy}} \bibnamefont{and}
  \bibinfo{author}{\bibfnamefont{G.~G.} \bibnamefont{Ryzhikh}},
  \bibinfo{journal}{J.~Phys.~B} \textbf{\bibinfo{volume}{32}},
  \bibinfo{pages}{2831} (\bibinfo{year}{1999}{\natexlab{a}}).

\bibitem[{\citenamefont{Bromley et~al.}(2000)\citenamefont{Bromley, Mitroy, and
  Ryzhikh}}]{bromley00a}
\bibinfo{author}{\bibfnamefont{M.~W.~J.} \bibnamefont{Bromley}},
  \bibinfo{author}{\bibfnamefont{J.}~\bibnamefont{Mitroy}}, \bibnamefont{and}
  \bibinfo{author}{\bibfnamefont{G.~G.} \bibnamefont{Ryzhikh}},
  \bibinfo{journal}{Nucl.~Instrum.~Methods~Phys.~Res.~B}
  \textbf{\bibinfo{volume}{171}}, \bibinfo{pages}{47} (\bibinfo{year}{2000}).

\bibitem[{\citenamefont{Mitroy}(1999)}]{mitroy99f}
\bibinfo{author}{\bibfnamefont{J.}~\bibnamefont{Mitroy}},
  \bibinfo{journal}{Aust.~J.~Phys.} \textbf{\bibinfo{volume}{52}},
  \bibinfo{pages}{973} (\bibinfo{year}{1999}).

\bibitem[{\citenamefont{Neamtan et~al.}(1962)\citenamefont{Neamtan, Darewych,
  and Oczkowski}}]{neamtan62}
\bibinfo{author}{\bibfnamefont{S.~M.} \bibnamefont{Neamtan}},
  \bibinfo{author}{\bibfnamefont{G.}~\bibnamefont{Darewych}}, \bibnamefont{and}
  \bibinfo{author}{\bibfnamefont{G.}~\bibnamefont{Oczkowski}},
  \bibinfo{journal}{Phys.~Rev.} \textbf{\bibinfo{volume}{126}},
  \bibinfo{pages}{193} (\bibinfo{year}{1962}).

\bibitem[{\citenamefont{Drachman}(1995)}]{drachman95b}
\bibinfo{author}{\bibfnamefont{R.~J.} \bibnamefont{Drachman}}, in
  \emph{\bibinfo{booktitle}{The Physics of Electronic and Atomic Collisions}},
  edited by \bibinfo{editor}{\bibfnamefont{L.~J.} \bibnamefont{Dube}},
  \bibinfo{editor}{\bibfnamefont{J.~B.~A.} \bibnamefont{Mitchell}},
  \bibinfo{editor}{\bibfnamefont{J.~W.} \bibnamefont{{McConkey}}},
  \bibnamefont{and} \bibinfo{editor}{\bibfnamefont{C.~E.} \bibnamefont{Brion}}
  (\bibinfo{publisher}{American Institute of Physics}, \bibinfo{address}{New
  York}, \bibinfo{year}{1995}), vol. \bibinfo{volume}{XIX}, p.
  \bibinfo{pages}{369}.

\bibitem[{\citenamefont{Ryzhikh and Mitroy}(1999)}]{ryzhikh99a}
\bibinfo{author}{\bibfnamefont{G.~G.} \bibnamefont{Ryzhikh}} \bibnamefont{and}
  \bibinfo{author}{\bibfnamefont{J.}~\bibnamefont{Mitroy}},
  \bibinfo{journal}{J.~Phys.~B} \textbf{\bibinfo{volume}{32}},
  \bibinfo{pages}{4051} (\bibinfo{year}{1999}).

\bibitem[{\citenamefont{Mitroy and Bromley}(2005)}]{mitroy05i}
\bibinfo{author}{\bibfnamefont{J.}~\bibnamefont{Mitroy}} \bibnamefont{and}
  \bibinfo{author}{\bibfnamefont{M.~W.~J.} \bibnamefont{Bromley}},
  \bibinfo{journal}{J.~Chem.~Phys.} \textbf{\bibinfo{volume}{123}},
  \bibinfo{pages}{017101} (\bibinfo{year}{2005}).

\bibitem[{\citenamefont{Mitroy and Bromley}(2006{\natexlab{a}})}]{mitroy06a}
\bibinfo{author}{\bibfnamefont{J.}~\bibnamefont{Mitroy}} \bibnamefont{and}
  \bibinfo{author}{\bibfnamefont{M.~W.~J.} \bibnamefont{Bromley}}, p.
  \bibinfo{pages}{in preparation} (\bibinfo{year}{2006}{\natexlab{a}}).

\bibitem[{\citenamefont{Mitroy and Bromley}(2006{\natexlab{b}})}]{mitroy06b}
\bibinfo{author}{\bibfnamefont{J.}~\bibnamefont{Mitroy}} \bibnamefont{and}
  \bibinfo{author}{\bibfnamefont{M.~W.~J.} \bibnamefont{Bromley}}, p.
  \bibinfo{pages}{under review} (\bibinfo{year}{2006}{\natexlab{b}}).

\bibitem[{\citenamefont{Gribakin and Ludlow}(2002)}]{gribakin02a}
\bibinfo{author}{\bibfnamefont{G.~F.} \bibnamefont{Gribakin}} \bibnamefont{and}
  \bibinfo{author}{\bibfnamefont{J.}~\bibnamefont{Ludlow}},
  \bibinfo{journal}{J.~Phys.~B} \textbf{\bibinfo{volume}{35}},
  \bibinfo{pages}{339} (\bibinfo{year}{2002}).

\bibitem[{\citenamefont{Schwartz}(1962)}]{schwartz62a}
\bibinfo{author}{\bibfnamefont{C.}~\bibnamefont{Schwartz}},
  \bibinfo{journal}{Phys.~Rev.} \textbf{\bibinfo{volume}{126}},
  \bibinfo{pages}{1015} (\bibinfo{year}{1962}).

\bibitem[{\citenamefont{Carroll et~al.}(1979)\citenamefont{Carroll,
  Silverstone, and Metzger}}]{carroll79a}
\bibinfo{author}{\bibfnamefont{D.~P.} \bibnamefont{Carroll}},
  \bibinfo{author}{\bibfnamefont{H.~J.} \bibnamefont{Silverstone}},
  \bibnamefont{and} \bibinfo{author}{\bibfnamefont{R.~P.}
  \bibnamefont{Metzger}}, \bibinfo{journal}{J.~Chem.~Phys.}
  \textbf{\bibinfo{volume}{71}}, \bibinfo{pages}{4142} (\bibinfo{year}{1979}).

\bibitem[{\citenamefont{Hill}(1985)}]{hill85a}
\bibinfo{author}{\bibfnamefont{R.~N.} \bibnamefont{Hill}},
  \bibinfo{journal}{J.~Chem.~Phys.} \textbf{\bibinfo{volume}{83}},
  \bibinfo{pages}{1173} (\bibinfo{year}{1985}).

\bibitem[{\citenamefont{Kutzelnigg and Morgan~III}(1992)}]{kutzelnigg92a}
\bibinfo{author}{\bibfnamefont{W.}~\bibnamefont{Kutzelnigg}} \bibnamefont{and}
  \bibinfo{author}{\bibfnamefont{J.~D.} \bibnamefont{Morgan~III}},
  \bibinfo{journal}{J.~Chem.~Phys.} \textbf{\bibinfo{volume}{96}},
  \bibinfo{pages}{4484} (\bibinfo{year}{1992}).

\bibitem[{\citenamefont{Schmidt and Linderberg}(1993)}]{schmidt93a}
\bibinfo{author}{\bibfnamefont{H.~M.} \bibnamefont{Schmidt}} \bibnamefont{and}
  \bibinfo{author}{\bibfnamefont{J.}~\bibnamefont{Linderberg}},
  \bibinfo{journal}{Phys.~Rev.~A} \textbf{\bibinfo{volume}{49}},
  \bibinfo{pages}{4404} (\bibinfo{year}{1993}).

\bibitem[{\citenamefont{Ottschofski and Kutzelnigg}(1997)}]{ottschofski97a}
\bibinfo{author}{\bibfnamefont{E.}~\bibnamefont{Ottschofski}} \bibnamefont{and}
  \bibinfo{author}{\bibfnamefont{W.}~\bibnamefont{Kutzelnigg}},
  \bibinfo{journal}{J.~Chem.~Phys.} \textbf{\bibinfo{volume}{106}},
  \bibinfo{pages}{6634} (\bibinfo{year}{1997}).

\bibitem[{\citenamefont{Bromley and Mitroy}(2006)}]{bromley06a}
\bibinfo{author}{\bibfnamefont{M.~W.~J.} \bibnamefont{Bromley}}
  \bibnamefont{and} \bibinfo{author}{\bibfnamefont{J.}~\bibnamefont{Mitroy}},
  \bibinfo{journal}{Int.~J.~Quantum~Chem.} p. \bibinfo{pages}{under review}
  (\bibinfo{year}{2006}).

\bibitem[{\citenamefont{Mitroy and Ryzhikh}(1999{\natexlab{b}})}]{mitroy99h}
\bibinfo{author}{\bibfnamefont{J.}~\bibnamefont{Mitroy}} \bibnamefont{and}
  \bibinfo{author}{\bibfnamefont{G.~G.} \bibnamefont{Ryzhikh}},
  \bibinfo{journal}{Comput.~Phys.~Commun.} \textbf{\bibinfo{volume}{123}},
  \bibinfo{pages}{103} (\bibinfo{year}{1999}{\natexlab{b}}).

\bibitem[{\citenamefont{Mitroy}(2005)}]{mitroy05d}
\bibinfo{author}{\bibfnamefont{J.}~\bibnamefont{Mitroy}},
  \bibinfo{journal}{Phys.~Rev.~Lett.} \textbf{\bibinfo{volume}{94}},
  \bibinfo{pages}{033402} (\bibinfo{year}{2005}).

\bibitem[{\citenamefont{{Krasnopol'sky} and Kukulin}(1974)}]{krasnopolsky74}
\bibinfo{author}{\bibfnamefont{V.~M.} \bibnamefont{{Krasnopol'sky}}}
  \bibnamefont{and} \bibinfo{author}{\bibfnamefont{V.~I.}
  \bibnamefont{Kukulin}}, \bibinfo{journal}{Sov.~J.~Nucl.~Phys.}
  \textbf{\bibinfo{volume}{20}}, \bibinfo{pages}{883} (\bibinfo{year}{1974}),
  \bibinfo{note}{yad.Fiz.(USSR) \textbf{20} (1974) 883}.

\bibitem[{\citenamefont{Stathopolous and {Froese
  Fischer}}(1994)}]{stathopolous94a}
\bibinfo{author}{\bibfnamefont{A.}~\bibnamefont{Stathopolous}}
  \bibnamefont{and} \bibinfo{author}{\bibfnamefont{C.}~\bibnamefont{{Froese
  Fischer}}}, \bibinfo{journal}{Comput.~Phys.~Commun.}
  \textbf{\bibinfo{volume}{79}}, \bibinfo{pages}{268} (\bibinfo{year}{1994}).

\bibitem[{\citenamefont{Mitroy et~al.}(1999)\citenamefont{Mitroy, Bromley, and
  Ryzhikh}}]{mitroy99b}
\bibinfo{author}{\bibfnamefont{J.}~\bibnamefont{Mitroy}},
  \bibinfo{author}{\bibfnamefont{M.~W.~J.} \bibnamefont{Bromley}},
  \bibnamefont{and} \bibinfo{author}{\bibfnamefont{G.~G.}
  \bibnamefont{Ryzhikh}}, \bibinfo{journal}{J.~Phys.~B}
  \textbf{\bibinfo{volume}{32}}, \bibinfo{pages}{2203} (\bibinfo{year}{1999}).

\bibitem[{\citenamefont{Mitroy and Novikov}(2004)}]{mitroy04a}
\bibinfo{author}{\bibfnamefont{J.}~\bibnamefont{Mitroy}} \bibnamefont{and}
  \bibinfo{author}{\bibfnamefont{S.~A.} \bibnamefont{Novikov}},
  \bibinfo{journal}{Phys.~Rev.~A} \textbf{\bibinfo{volume}{70}},
  \bibinfo{pages}{032511} (\bibinfo{year}{2004}).

\bibitem[{\citenamefont{Mella et~al.}(2002)\citenamefont{Mella, Casalegno, and
  Morosi}}]{mella02a}
\bibinfo{author}{\bibfnamefont{M.}~\bibnamefont{Mella}},
  \bibinfo{author}{\bibfnamefont{M.}~\bibnamefont{Casalegno}},
  \bibnamefont{and} \bibinfo{author}{\bibfnamefont{G.}~\bibnamefont{Morosi}},
  \bibinfo{journal}{J.~Chem.~Phys} \textbf{\bibinfo{volume}{117}},
  \bibinfo{pages}{1450} (\bibinfo{year}{2002}).

\bibitem[{\citenamefont{Szmytkowski}(1993)}]{szmytkowski93a}
\bibinfo{author}{\bibfnamefont{R.}~\bibnamefont{Szmytkowski}},
  \bibinfo{journal}{J.~Phys.~II} \textbf{\bibinfo{volume}{3}},
  \bibinfo{pages}{183} (\bibinfo{year}{1993}).

\bibitem[{\citenamefont{McEachran and Stauffer}(1998)}]{mceachran98}
\bibinfo{author}{\bibfnamefont{R.}~\bibnamefont{McEachran}} \bibnamefont{and}
  \bibinfo{author}{\bibfnamefont{A.~D.} \bibnamefont{Stauffer}},
  \bibinfo{journal}{Nucl.~Instrum.~Methods~Phys.~Res.~B}
  \textbf{\bibinfo{volume}{143}}, \bibinfo{pages}{199} (\bibinfo{year}{1998}).

\bibitem[{\citenamefont{Gribakin and King}(1996)}]{gribakin96}
\bibinfo{author}{\bibfnamefont{G.~F.} \bibnamefont{Gribakin}} \bibnamefont{and}
  \bibinfo{author}{\bibfnamefont{W.~A.} \bibnamefont{King}},
  \bibinfo{journal}{Can.~J.~Phys.} \textbf{\bibinfo{volume}{74}},
  \bibinfo{pages}{449} (\bibinfo{year}{1996}).

\bibitem[{\citenamefont{Mitroy and Bromley}(2006{\natexlab{c}})}]{mitroy06c}
\bibinfo{author}{\bibfnamefont{J.}~\bibnamefont{Mitroy}} \bibnamefont{and}
  \bibinfo{author}{\bibfnamefont{M.~W.~J.} \bibnamefont{Bromley}}, p.
  \bibinfo{pages}{in preparation} (\bibinfo{year}{2006}{\natexlab{c}}).

\bibitem[{\citenamefont{Whitehead et~al.}(1977)\citenamefont{Whitehead, Watt,
  Cole, and Morrison}}]{whitehead77a}
\bibinfo{author}{\bibfnamefont{R.~R.} \bibnamefont{Whitehead}},
  \bibinfo{author}{\bibfnamefont{A.}~\bibnamefont{Watt}},
  \bibinfo{author}{\bibfnamefont{B.~J.} \bibnamefont{Cole}}, \bibnamefont{and}
  \bibinfo{author}{\bibfnamefont{I.}~\bibnamefont{Morrison}},
  \bibinfo{journal}{Adv.~Nucl.~Phys.} \textbf{\bibinfo{volume}{9}},
  \bibinfo{pages}{123} (\bibinfo{year}{1977}).

\end{thebibliography}

\end{document}